\documentclass[twocolumn,showpacs,amssymb,prd]{revtex4}
\begin{document}

\def\btt#1{{\tt$\backslash$#1}}




\newcommand{\gsim}{\hbox{\rlap{$^>$}$_\sim$}}
\newcommand{\lsim}{\hbox{\rlap{$^<$}$_\sim$}}

 \title{Possible Origin Of The Neutrino Speed Anomaly\\ 
               Reported By OPERA}

\author{Shlomo Dado and Arnon Dar} \affiliation{Department of Physics, 
Technion,
Haifa 32000, Israel}

\begin{abstract}

Recently the OPERA collaboration reported a measurement of a 
superluminal speed of muon neutrinos traveling through the Earth's 
crust between their production site at CERN and their detection site 
under Gran Sasso, $\!\sim\!730$ km away. The measurement was based on 
the assumption that the pulse shape of the neutrinos from the decay of 
parent mesons produced in proton-target collisions is the 
same as that of the incident protons.  Here we argue that the 
effective column density of the target along the beam direction 
decreases with time during the 10.5 $\mu$s duration of the proton 
pulse. This is because of the thermal expansion and expulsion of 
target material along the beam by the energy-momentum deposition 
during the 10.5 $\mu$s pulse. The progressive reduction in the 
effective column density during the pulse decreases the neutrino 
production rate per incident proton. It could   
have advanced the mean production time of the detected neutrinos 
relative to that calculated from the proton pulse-shape, by an amount 
comparable to the measured neutrino lead time ${\rm 60.7 \pm 6.9(stat) 
\pm 7.4 (sys)}$ ns. This explanation implies that the planned 
measurements by OPERA of the speed of neutrinos produced in much 
shorter (a few ns) pulses, should yield a speed consistent with the 
speed of light in free space.

\end{abstract} 

\maketitle
\section{Introduction}

The OPERA collaboration has reported recently \cite{OPERA}
a measurement of a superluminal speed
${\rm (v/c-1)=[2.48\pm 0.28(stat)\pm 0.30(sys)]\times 10^{-5}}$ of
muon neutrinos traveling through earth between their production site
at CERN and their detection site under Gran Sasso, $\!\sim\!730$ km
away. However, the OPERA measurement of the neutrino speed 
was not based on the time difference between the production and 
detection of individual neutrinos but rather on the 
measured distribution of arrival times of 
neutrinos at Gran Sasso  with respect to the 
waveform of the pulsed protons entering the CERN Neutrino to Gran Sasso 
(CNGS) system.  
Although the distance $d$ from the target where the parent meson produces 
the neutrino in 
the decay tunnel is unknown, this introduces negligible difference 
between the time that a proton enters the CNGS system and the time  
that the produced neutrino enters 
the OPERA detector under the Gran Sasso,
\begin{equation}
dt\approx d/2c\gamma_m^2\sim \tau_m/2\gamma_m \lsim 0.1\, {\rm ns}  
\label{pidecay} 
\end{equation}
for pions of lifetime $\tau_m=\tau_\pi\approx 2\times 10^{-8}$ s 
and Lorentz factor $\gamma_m=\gamma_\pi\!\geq\!100$.    

The neutrino speed that was inferred by OPERA
was based on the assumption that the probability 
to produce the parent meson in the target is the same for all 
the protons in the pulse, and  hence, the temporal  profile of 
the neutrino pulse is  faithfully represented   
by temporal profile of the proton pulse. 
However, this may not be the case because of 
the large energy-momentum deposition  in the graphite target 
along the beam during the 10.5 $\mu$s proton pulse.
This large  energy-momentum deposition in the target,
as we shall argue below, probably
reduces  the meson production rate for late incident 
protons during a pulse 
due to thermal expansion and expulsion of target material  from the 
beam path. This can advance the  mean production time 
of the neutrinos relative to the mean time  
of the proton pulse by the neutrino 60.7 ns lead time that was 
inferred  by the OPERA collaboration.

\section{The neutrino production rate per incident proton}

Let us first demonstrate that a modest decrease in the neutrino 
production rate per incident proton between the beginning and the end 
of the proton pulse can explain the OPERA anomaly.
The proton pulse shape  has a leading edge rise-time of about 0.8 
$\mu$s  and a trailing edge fall-time of
about 0.4 $\mu$s  and approximately a flat top during the 9.3 $\mu$s
in between (see Fig. 11 of\cite{OPERA}). 
For simplicity, let us approximate the proton pulse shape
by a rectangular pulse of  duration  $\Delta t\!=\!10.5\, 
\mu$s.
Let us also assume that due to the change in the effective 
column density in the graphite target along the beam direction,
the neutrino production rate per incident proton decreases linearly  
with time during the pulse. Consequently, the neutrino pulse shape  
becomes    
$N(t)=N(0)\!-\!(\Delta N/\Delta t)\,t$  where  
$\Delta N= N(0)-N(\Delta t)$. 

For a small change, $\Delta N\! \ll\! N(0) $,  the mean time of such 
a neutrino pulse is given by,
\begin{equation}
t_m(\nu) \simeq {\Delta t\over 2}\,
\left(1-{\Delta N\over 6\,N(0)}\right)\, . 
\label{tm} 
\end{equation}
The observed lead time of 60.7 ns in the mean arrival time of the 
neutrinos requires that the effective column density 
encountered by the proton beam decreases by 
$\approx 7\% $ during the proton pulse.

\section{The target respond  to the beam}

It would be presumptuous to 
calculate in detail the changes that take place 
along the beam path in the CNGS target 
during the 10.5 $\mu$s proton pulse. Thus, we shall limit ourselves
to simple estimates.       

In the OPERA experiment the beam  consists  of 
400 GeV protons extracted from the CERN Super
Proton Synchrotron (SPS). 
The beam cycle is typically 6 s.
There are two extractions per cycle,
each 10.5 $\mu$s, separated by 50 ms. 
Each extraction has $2.4 \times 10^{13}$ protons,
which carry  $\simeq 1.5\times 10^{13}$ erg. 
The beam is nearly cylindrical with a diameter of 0.5 mm.

The target is 2 m long sealed container,  which contains 13 
graphite rods 100 mm long each.
The first two have  5 mm diameter and the rest 
have 4 mm  diameter. The total weight of the graphite is 
$\!\sim\! 41$ g  (for specific density of 2.3 g cm$^{-3}$).
The specific heat of graphite is 0.17 calories/g$^o$K,
i.e.,  $7.1\times 10^6\, {\rm erg/ g\, ^oK}$. A very large number of 
disks  outside the container cool the target by heat exchange with 
the room temperature during the 6 s cycle.

Most of the prompt energy loss is through the escape of
energetic particles which are produced in the proton initiated showers 
within the target. 
However  a significant  fraction  of the energy deposited in the 
target,  ($\sim 10^{12}$ erg)  does not escape during 
the 10.5 $\mu$s pulse 
and heats the target to a very high temperature.
The rising temperature 
decreases the target density, in particular along the beam direction.
Well below the Graphite melting temperature ($\sim 3650\, ^o$K), 
the thermal linear expansion coefficient 
of graphite is  $\sim 7.9\times 10^{-6}\,{\rm {(^oK)}^{-1}} $ yielding  
$\sim 4\%$ column density decrease at roughly 2500$^O$K.
Although,  the thermal expansion coefficient of the target  
at such high temperatures  is not well known,  a reduction 
of 7\% (see section 2) in the effective column density of the graphite 
target along the beam path at the end of the proton pulse seems plausible.
The thermal radiation from the container and the disks cool the target  
during the 6 s cycle back to its  much lower initial temperature.

\section{Conclusions}

It has been noticed before (e.g. \cite{Knobloch2011}), that the assumption 
of the same temporal profile of the neutrino pulse as that of the proton 
pulse could be responsible for the neutrino speed anomaly, which was 
measured by the OPERA collaboration. In this short note we have proposed a 
plausible origin for a difference between the proton and neutrino pulse 
shapes and we showed that it could have led to the neutrino speed anomaly 
reported by OPERA. In particular, we predict that once the extracted 10.5 
$\mu$s proton pulses in the CNGS will be replaced by a few ns pulses the 
neutrino speed anomaly will disappear.

\noindent
{\bf Acknowledgment:}
We would like to thank Jacques Goldberg  
for providing us with useful details on the OPERA experiment.

{}
\end{document}